# Lifetime measurements of nuclei in few-electron ions

Thomas Faestermann

Faculty of Physics E12, Technische Universität München and Excellence Cluster Universe, D85748 Garching, Germany

thomas.faestermann@mytum.de

**Abstract**
In this review lifetime measurements of ions with at most two electrons are summarized. Such highly ionized systems have been studied - until now - only in the Experimental Storage Ring of the GSI in Darmstadt. Emphasis is put on decays via the weak interaction. The first observations of beta-decay into bound atomic states are described as well as its time mirrored counterpart, the electron-capture decay. In the latter case the decays of hydrogen- and helium-like ions are compared with a surprising result. Further on, the observation of sinusoidal modulations of the decay rate in two-body decays is summarized. As a possible cause an interference due to the emission of neutrinos with different rest mass is discussed.

## 1. Introduction: the ESR

Until recently, the Experimental Storage Ring (ESR) of the GSI Helmholtz Centre for Heavy Ion Physics was the only installation where highly charged (even bare) heavy ions could be stored and investigated. It was designed by Bernhard Franzke and his group (Franzke 1987). Funding was applied for it together with the synchrotron SIS18 and approved under the GSI directorate of the late Paul Kienle, who promoted physics and especially nuclear physics experiments at the ESR until the end of his life in 2013. The ESR went into operation in 1990. The main characteristics are: a circumference of 108.36 m; a maximum rigidity of 10 Tm; this corresponds for bare N=Z nuclei to 830 MeV/u. It is provided with electron cooling (Steck *et al* 2004) with a maximum electron energy of 310 keV corresponding to ions with 564 MeV/u. With a total number of less than about 2000 ions in the ESR a relative momentum spread of $<10^{-6}$ can be achieved (Steck *et al* 1996). In order to precool secondary ions produced by fragmentation reactions and injected through the fragment separator FRS also stochastic cooling has been installed by Nolden *et al* 2004. This operates typically at a velocity β=0.71 which corresponds to 400 MeV/u.

The revolution frequency of ions can be measured by a Fourier transform of the so-called Schottky noise induced in capacitive pick-up plates. Typically a high (about 30$^{th}$) harmonic of the revolution frequency of about 2 MHz is used. This Schottky noise information is sensitive enough to observe single highly charged ions after a measuring period of a few 100 ms and to separate ions with a mass difference of a few 100 keV/$c^2$.

To increase the sensitivity, recently a resonant pick-up was constructed and successfully installed (Sanjari *et al* 2013). It has a resonance frequency of 245 MHz and typically the 124$^{th}$ harmonic of the revolution frequency is used for stochastically cooled ions. The high quality factor of Q > 1000 and the higher frequency result in much higher sensitivity at shorter sampling times of a few 10 ms.

In addition to these non-destructive detection methods also particle detectors of different kind have been used especially to detect ions which have changed their magnetic rigidity by stripping of an electron in the gas target or by decays that change the mass and/or the charge.

All this equipment has been used for atomic physics experiments, but also for measuring nuclear masses and lifetimes. In this review the main emphasis is put on nuclear beta-decay lifetimes.

## 2. Bound-state beta-decay

In principle, it is always possible in a β⁻-decay, that the decay electron gets bound in an atomic orbit and only an antineutrino is emitted. However the probability for that process is extremely small, if the strongly bound atomic levels of the decaying nucleus are all filled. But if the nucleus has none or only one electron, the decay electron can stay in the K-shell and the phase space is large. Therefore this process of bound-state β-decay could for the first time be observed at the ESR for bare $^{163}$Dy nuclei by M. Jung *et al* (1992). Neutral $^{163}$Dy atoms are stable, whereas $^{163}$Ho atoms decay by electron-capture to $^{163}$Dy with a Q-value of 2.6 keV. If all electrons are removed, $^{163}$Dy has even a more negative Q-value for continuums β-decay, but if the decay-electron remains in the K-shell, the Q-value is +50 keV. In the experiment bare $^{163}$Dy$^{66+}$ ions from the SIS18 were injected into the ESR and monitored with the Schottky diagnosis. The $^{163}$Ho$^{66+}$ ions from the bound β-decay could not directly be seen since the charge is not changed and the mass only by 50 keV/c$^2$. Therefore, after some waiting time in the order of hours, the gas jet target was turned on which stripped the electron off the hydrogen-like $^{163}$Ho$^{66+}$ ions which had been created by bound-state β-decay. The $^{163}$Ho$^{67+}$ ions were deflected more in the following dipole and detected with a multi-wire proportional counter. In this way the ratio of decayed ions to the total number was measured as a function of the waiting time and thus the half-life of bare $^{163}$Dy$^{66+}$ nuclei as $T_{1/2} = (48^{+5}_{-4})$ days. This information has been used to derive the temperature in the s-process branching from $^{163}$Dy towards $^{164}$Er as kT=(28±8) keV (Bosch 2008).

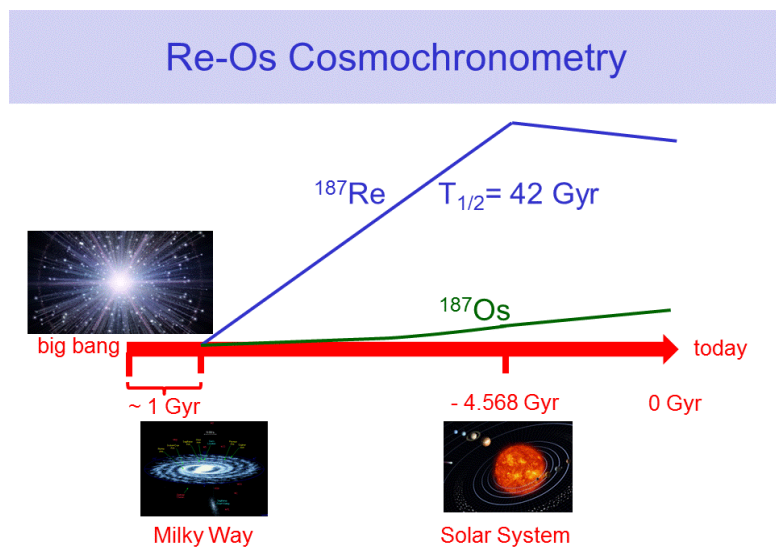

Figure 1: Schematic of Re-Os cosmochronometry. From the formation of the Galaxy on $^{187}$Re is produced through the r-process. The $^{187}$Os content grows - besides s-process production – from the β-decay of $^{187}$Re. The relative concentration today in meteorites can be followed back to the birth of the Solar system and thus yields an estimate for the time during which r-process in the Galaxy took place.

Another astrophysically interesting pair of isotopes is the pair $^{187}$Re and $^{187}$Os. Neutral $^{187}$Re has a half-life of 42 Gyr, by far exceeding the age of the universe. Therefore the relative abundance can be used as a cosmochronometer for the age of our Galaxy. In the r-process $^{187}$Re is produced in the β-decay of more neutron rich A=187 isotopes, but $^{187}$Os is shielded. It builds up from the β-decay of $^{187}$Re. The rough idea is sketched in figure 1, where a constant star formation and r-process rate is assumed enriching the interstellar matter with $^{187}$Re and through decay $^{187}$Os until the Solar system is formed from this material, 4.568 Gyr ago. Afterwards the decay of $^{187}$Re and the increase of $^{187}$Os until today can easily be calculated. Thus from the Solar abundances of these isotopes today the duration of nucleosynthesis in the Galaxy and thus the age of the Galaxy and the age of the universe may be estimated. One of the uncertainties is the modification that occurs if $^{187}$Re gets into a hot environment and decays faster because it is highly ionized and bound-state β-decay becomes the dominant decay channel. Since the latter decay mode can proceed to an excited state at 9.75 keV in $^{187}$Os, the nuclear matrix element is not known. Therefore bound-state β-decay of bare $^{187}$Re$^{75+}$ has been measured in the ESR by Bosch *et al* in 1996. The procedure was similar as for $^{163}$Dy. In this case a specially built micro-strip gas counter was used to detect the totally stripped $^{187}$Os$^{76+}$ daughter ions after the waiting time. As an alternative for half of the measuring time the number of stripped daughter ions was measured with the Schottky analysis. This was possible since the ESR was tuned in such a way that both charge states 75+ and 76+ were coasting in the ESR and could now be distinguished. The results of both methods agreed and yielded a half-life for bare $^{187}$Re$^{75+}$ of $T_{1/2}$ = (32.9±2.0) *yr*. This is more than 9 orders of magnitude smaller than that of neutral $^{187}$Re. About equal factors are due to the larger Q-value and due to the larger nuclear matrix element to the first excited state in $^{187}$Os, which is a simple first forbidden transition, whereas that to the ground state is unique first forbidden.

Takahashi *et al* (1998a,1998b and 2003) used this measured log(ft) value to estimate the age of the Galaxy to $T_G = (15^{+2}_{-3})Gyr$ which at that time was similar as the astronomical ages derived from globular clusters (e.g. $T_G = (14^{+3.5}_{-3.0})Gyr$ with 2 σ uncertainty, Chaboyer *et al* 1996) but still consistent with the most recent CMB age of the Universe $T_U = (13.799 \pm 0.038)Gyr$ (Adam 2015), even if 1 Gyr is subtracted for the formation of our Galaxy. He started from the meteoritic $^{187}$Re and $^{186,187}$Os abundances (Anders and Grevesse 1989, Faestermann 1998) and took into account different models of the chemical evolution and the mass distribution of stars being born in our Galaxy. The influence of neutron-capture cross sections of the Os isotopes on the s-process production and depletion has been studied by Mosconi *et al* (2007) and Mengoni *et al* (2009).

Bound-state β-decay was for the first time directly observed at the ESR for $^{207}$Tl by Ohtsubo *et al* (2005). The bare $^{207}$Tl$^{81+}$ mother ions were well separated from the bound-state β-decay daughter $^{207}$Pb$^{81+}$ (atomic mass difference Δm = 1418 keV/c$^2$) in the Schottky analysis. From the decrease of the $^{207}$Tl$^{81+}$ mother ions with time the total decay constant $\lambda_{tot}=\lambda_{cont}+\lambda_{bound}+\lambda_{loss}$, and from the grow-in of the $^{207}$Pb$^{81+}$ daughter ions the partial decay constant for bound-state beta decay $\lambda_{bound}$ could be determined. The decay constant for losses due to charge changing collisions $\lambda_{loss}$ was determined separately with stable ions. Although the continuum β-decay of bare $^{207}$Tl in its rest frame is slower than for neutral atoms, the total decay constant is larger due to the bound β-decay, which contributes about 16% to the summed decay rate. In addition to this Schottky analysis the daughter ions from continuum β-decay were intercepted by a silicon detector telescope that measured position and energy loss of ions deflected more than the coasting ions. From these measurements consistent values for the total decay rate were deduced (Maier *et al* 2006).

## 3. Electron capture decay

The better known two-body decay process in weak interactions is electron-capture (EC) decay where only a neutrino is emitted. This process has also been studied for few-electron systems at the ESR. The first case was the decay of the $1^+$ ground state of $^{140}$Pr (Litvinov *et al* 2007). As a neutral atom it decays with 99.4% to the $0^+$ ground state $^{140}$Ce and the $\beta^+$-decay Q-value is 3388 keV. About half of the decays is via EC. Now the question arises how the decay probability for EC changes when only one or two electrons are in the atomic shell, i.e. for H-like or He-like ions. Again $^{140}$Pr ions in the 58+ or 57+ charge state were produced by fragmentation and injected into the ESR and the number of coasting ions, well distinguished mother and daughter ions in the same charge state were followed by Schottky spectroscopy as a function of time. The surprizing result was that the nucleus with one electron, $^{140}$Pr$^{58+}$, decays faster via EC than that with two electrons, $^{140}$Pr$^{57+}$. The measured ratio of the decay constants is $\lambda_{EC}(^{140}\text{Pr}^{58+})/\lambda_{EC}(^{140}\text{Pr}^{57+}) = 1.49\pm0.06$. The explanation lies in the conservation of the total angular momentum $\vec{F} = \vec{I} + \vec{S}$ where I and S are nuclear and atomic spin respectively. The H-like $^{140}$Pr$^{58+}$ with one electron in the K-shell can have F=3/2 or F=1/2, but in the final state the bare $^{140}$Ce$^{58+}$ is spin-less and only the emitted neutrino has spin 1/2, thus the decay of the F=3/2 state is forbidden. If in $^{140}$Pr$^{58+}$ all hyperfine states were equally populated, i.e. with the weights (2F+1) or 2/3 and 1/3, the decay rate would indeed be half of that for the two-electron ion $^{140}$Pr$^{57+}$. But, because of the positive magnetic moment of $^{140}$Pr the F=1/2 state is lower in energy and populated completely after short time, a factor of 3 is gained and the expected ratio is 3/2 as measured. In a more rigorous treatment the decay rates have been calculated by Patyk *et al* 2008, Ivanov *et al* 2008 and Siegien-Iwaniuk *et al* 2012.

Similar measurements with the same result have been done for H-like and He-like $^{142}$Pm and $^{122}$I ions which have also the same nuclear spin and parity $1^+$ of the ground state (Winckler *et al* 2009a, Atanasov *et al* 2012).

## 4. Decay rate modulations

In two-body weak decays, like EC decay, a monoenergetic neutrino is emitted. Since it is well established, that neutrinos are massive, and that the weak eigenstates are linear combinations of the mass eigenstates, the question arose, whether the possibility to emit neutrinos of different mass could lead to interferences in the decay probability. Therefore a program was started to investigate the EC decays described in section 3 in more detail. To observe changes in the decay rate, it is not sufficient to measure the number of mother and daughter ions as a function of time. One has to observe every individual decay time. And that requires, with the sensitivity of the capacitive Schottky analysis, that only few mother ions are injected into the ESR and not more than two EC-decays occur in the observation time. A time consuming task indeed! The first systems studied were H-like $^{140}$Pr$^{58+}$ and $^{142}$Pm$^{60+}$ (Litvinov *et al* 2008). Many thousand injections and measuring cycles of nearly 2 min duration were necessary to achieve reasonable statistics. The times when a mother ion disappeared and, after electron-cooling, the EC-daughter ion appeared in the frequency spectra were visually determined. The distribution of decay times was then binned and decay time spectra were generated. These were fitted either with purely exponential decays, or with a sinusoidal modulation on top of the exponential and three additional parameters: amplitude, frequency and phase. In both cases the fit with modulation was favoured by the resulting $\chi^2$. The fit results are given in table 1. The resulting frequencies for the two systems are the same within 0.4%. To investigate whether there is a dependence on the mass of the ions, in the same way the decay of H-like $^{122}$I$^{52+}$ was measured and also a significant modulation was observed. Unfortunately, since the collaboration was waiting for an unbiased automatic analysis, there is no final publication of these data. But results of the preliminary manual analysis (see figure 2) are published (Winckler *et al* 2009b, Kienle 2009). Table 1 gives the most recent, still not final numbers.

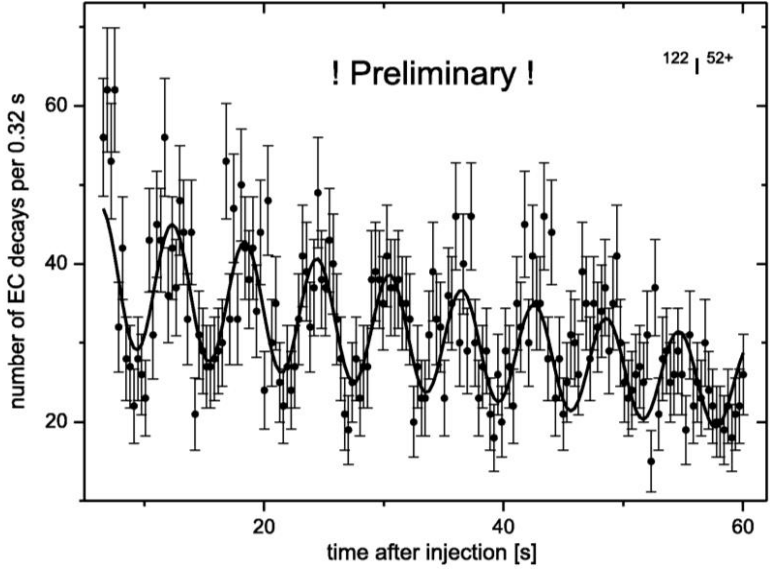

Figure 2: Preliminary results for EC-decay of H-like $^{122}I^{52+}$ ions (Winckler *et al* 2009b) .

Immediately the question came up, whether such a modulation was overlooked in earlier experiments measuring the EC-decay of atoms in a solid or fluid environment or whether the modulation is specific to the experiments with freely moving H-like ions. With that question in mind experiments have been done on radioactive nuclei in neutral atoms: on $^{142}$Pm (Vetter *et al* 2008) on $^{180}$Re (Faestermann *et al* 2009) and on $^{140}$Pr (Yamaguchi *et al* 2014). In the first and third case the K x-rays after EC were used to yield the time of decay, in the second case the EC-decay proceeds to an excited state in $^{180}$W and the following γ-ray was used as trigger. In none of the cases a modulation of the decay curve was observed, indicating that either the long lived final state in the ESR experiments or the negligible influence from neighbouring ions might be essential for the modulation.

The new resonant Schottky pick-up described in Sec. 1 offered the possibility to repeat a measurement with much higher sensitivity. The case of $^{142}$Pm was chosen and the EC-decay times were measured simultaneously with both systems (Kienle *et al* 2013). Part of the data in this experiment could not be used for the analysis, since the injection kicker, also responsible for removing the ions in the ESR before a new run was started, did not work properly. The results both for the capacitive and the resonant pick-up (table 1) agree and show a modulation with the same frequency as before, however with a smaller amplitude.

The most naïve way, how one could think of interference to happen, is to assume that the different recoil energies which the daughter ions would assume when emitting neutrinos of different mass could cause interference. Restricting ourselves to two neutrino species *j=1,2* we can calculate the difference in recoil energies $\Delta E_R$ (which equals the difference in neutrino energies) from energy conservation $M_P c^2 = E_D + E_j$, square it and use momentum conservation $p_D = p_j$ (equation (1) is to be read from the second equals sign outward):

$$M_P^2 c^4 + E_j^2 - 2 M_P c^2 E_j = (M_P c^2 - E_j)^2 = E_D^2 = M_D^2 c^4 + p_D^2 c^2 = M_D^2 c^4 + E_j^2 - m_j^2 c^4 \quad (1)$$

subtracting (1) for j=1,2:  $\Delta m^2 = m_2^2 - m_1^2 = 2M_P(E_2 - E_1)/c^2 = 2M_P \Delta E_R / c^2$     (2)

with $p_{1,2}$ and $m_{1,2}$ the momenta and masses of neutrino species 1 and 2, $M_P$, $M_D$, $E_P$ and $E_D$ the mass and energy of parent and recoiling daughter ion and $E_1$, $E_2$ the total neutrino energies which differ only by $\Delta E_R$. This might lead to an oscillation with $\omega = \Delta E_R / \hbar$. That means that $\omega$ would be related to the neutrino mass difference $\Delta m^2 = m_2^2 - m_1^2$ via $\Delta m^2 c^4 = 2\hbar\omega\, M_P c^2$. And since we measure $\omega_{lab}$ in the laboratory frame we have to transform into the moving system with $\omega = \gamma\omega_{lab}$. The quantity $2\hbar\omega_{lab}\gamma M_P c^2$ is also listed in table 1. It is surprizing that the values of this expression agree for four independent measurements, one with two methods, because the dependence on the mass of the ions seems well reproduced. However, the value of this expression is a factor of 3 larger than the observed value from mainly SNO and KAMLAND $\Delta m_{2,1}^2 c^4 = (0.753 \pm 0.018) \times 10^{-4} eV^2$ (Olive et al 2014).

Anyway, this naïve picture is heavily criticized by practically all of the many theoreticians working on this problem (see references in Kienle et al 2013). There is only one group (Ivanov and Kienle 2010, 2014), which theoretically explains all observations. And, it should be mentioned, that a recent publication (Potzel 2014) explains the missing factor of 3 with the relativistics of the circulating ions and deduces from the modulation frequencies $\Delta m_{2,1}^2 c^4 = (0.768 \pm 0.012) \times 10^{-4} eV^2$ !

Table 1. Fitted oscillation parameters for the four experiments. Values for $^{122}$I are preliminary. The last two rows are for the capacitive and resonant pick-up.

| nucleus | amplitude | $\omega$ [s$^{-1}$] | $2\hbar\omega\gamma Mc^2$ [$10^{-4}$eV$^2$] | Reference |
|---|---|---|---|---|
| $^{140}$Pr | 0.18(3) | 0.890(11) | 2.182(27) | Litvinov 2008 |
| $^{142}$Pm | 0.23(4) | 0.885(31) | 2.20(8) | Litvinov 2008 |
| $^{122}$I | 0.176(28) | 1.014(10) | 2.164(20) | to be published |
| $^{142}$Pm | 0.134(27) | 0.882(14) | 2.194(35) | Kienle 2013 |
| $^{142}$Pm | 0.107(24) | 0.884(14) | 2.199(35) | Kienle 2013 |

5. **Outlook**

Since the observation and interpretation of these decay rate modulations is highly controversial, it is desirable that more such measurements are performed, also at independent institutions. The only other storage ring which should be able to perform such measurements in the near future is the CSRe storage ring at the Institute of Modern Physics in Lanzhou, China (e.g. Zhan et al 2010) which is very similar to the ESR.

On the other hand also Penning or Paul traps could be used for the observation of EC-decays. Here also the influence of neighbouring atoms or ions is very small. However, ions in traps necessarily have small energies in the order of eV and even if one injects highly charged ions one needs extremely good vacuum to keep the charge state. According to the Bohr criterion, those atomic electrons are stripped off an ion which collides with an electron gas, whose orbital velocity is in the same range as the relative velocity of ion and electron gas. The Bohr velocity of electrons in the K-shell is $v_0 = Z\alpha \cdot c$. Thus one should look for cases with low Z, in order to facilitate stripping into the K-shell. On the other hand the probability of EC-decay scales with the density of K-electrons at the nucleus, thus with $Z^3$ and calls for high Z. To increase the EC rate $\lambda_{EC}$ one can look for decays with small Q-value, but then the total decay rate gets small. And the total half-life should not be too long compared to the "expected" period of the oscillation, ideally 10-20 times as long. So, the conflicting requirements for a good case are: low Z, low Q, large $\lambda_{EC}$, small $T_{1/2}$, large EC fraction. In table 2 the candidates are listed up to about A=100, the "expected" period of the oscillation is the centre of mass period of $^{140}$Pr scaled with A.

Apparently there is no ideal case. In my judgement the case of $^{19}$Ne could be the most promising. It can be produced even at a small accelerator with direct reactions. It should not be hard to strip off 8 or 9 electrons, since the Bohr velocity for K-electrons is $\beta=0.07$. The half-life of 17 s is convenient and 26 times the oscillation period. For the H-like ion we have to check whether the lowest hyperfine state can decay via EC because of angular momentum conservation. Since $^{19}$Ne is an odd neutron nucleus it has for the $1/2^+$ ground state a negative magnetic moment of $\mu = -1.9\mu_N$. Thus, in the H-like ion the F=1 state is the atomic ground state. It can decay to the mirror ground state of $^{19}$F$^{9+}$, if the neutrino spin is parallel to the nuclear spin. The He-like ion has no hyperfine splitting anyway. A problem is, that only one in 1000 decays proceeds via EC.

Similarly as in the ESR the daughter ion could be detected in a Penning trap via the mirror charges induced by the cyclotron motion in the trap electrodes with the method of Fourier transform ion cyclotron resonance (FT-ICR, Schabinger *et al* 2012). The time necessary for detection is at the moment in the order of a few 100 ms, a little long for an oscillation period of 679 ms, but could be made much shorter (Blaum 2015), since the required resolution is not high for the relative change of cyclotron frequencies of $2\times10^{-4}$ between $^{19}$Ne$^{9+}$ and $^{19}$F$^{9+}$.

**Acknowledgements**

I am grateful to many people, especially to my old friend Fritz Bosch, who proposed most of the described experiments, and to my teacher Paul Kienle, who got me involved in this, and many other subjects. These experiments became only possible due to the art of the ESR-sorcerers: Bernhard Franzke, Markus Steck, Fritz Nolden, Shahab Sanjari and their teams. The very important interface between FRS and ESR was formed by Yuri Litvinov, and – not only in these experiments – the FRS masters Hans Geissel and Helmut Weick helped me a great deal. This work was partly funded by the DFG Cluster of Excellence "Origin and Structure of the Universe".

**Table 2.** Parameters to find suitable EC emitters for measurements in traps.

| nucleus | $T_{1/2}$ | unit | $\lambda$ [s$^{-1}$] | %EC | $\lambda$(EC) [s$^{-1}$] | Period s | $T_{1/2}$/Period |
|---|---|---|---|---|---|---|---|
| $^{7}$Be | 53.29 | d | 1.51E-07 | 100% | 1.51E-07 | 0.250 | 1.84E+07 |
| $^{11}$C | 20.38 | min | 5.67E-04 | 0.233% | 1.32E-06 | 0.393 | 3.11E+03 |
| $^{13}$N | 9.96 | min | 1.16E-03 | 0.196% | 2.28E-06 | 0.464 | 1.29E+03 |
| $^{15}$O | 2.03 | min | 5.69E-03 | 0.100% | 5.67E-06 | 0.536 | 2.27E+02 |
| $^{17}$F | 64.8 | min | 1.78E-04 | 0.15% | 2.60E-07 | 0.607 | 6.40E+03 |
| $^{18}$F | 109.77 | min | 1.05E-04 | 3.27% | 3.44E-06 | 0.643 | 1.02E+04 |
| $^{18}$Ne | 1.672 | s | 4.15E-01 | 0.025% | 1.04E-04 | 0.643 | 2.60E+00 |
| $^{19}$Ne | 17.22 | s | 4.03E-02 | 0.10% | 4.03E-05 | 0.679 | 2.54E+01 |
| $^{25}$Al | 7.18 | s | 9.65E-02 | 0.08% | 7.57E-05 | 0.893 | 8.04E+00 |
| $^{26}$Al$^{m}$ | 6.3502 | s | 1.09E-01 | 0.082% | 8.95E-05 | 0.929 | 6.84E+00 |
| $^{37}$Ar | 35 | d | 2.29E-07 | 100% | 2.29E-07 | 1.321 | 2.29E+06 |
| $^{64}$Cu | 12.7 | h | 1.52E-05 | 43.20% | 6.55E-06 | 2.286 | 2.00E+04 |
| $^{68}$Ga | 67.63 | min | 1.71E-04 | 8.92% | 1.52E-05 | 2.429 | 1.67E+03 |
| $^{78}$Br | 6.46 | min | 1.79E-03 | 5.49% | 9.82E-05 | 2.786 | 1.39E+02 |